\documentstyle[12pt]{article}

\begin{document}

\hoffset = -0.3truecm
\voffset = -1.1truecm

\title{\bf
HALF-MONOPOLE AND MULTIMONOPOLE\footnote{USM Preprint, June 2004; Accepted for publication in the International Journal of Modern Physics A}}

\author{
{\bf Rosy Teh\footnote{email address: rosyteh@usm.my} and K. M. Wong}\\
{\normalsize School of Physics, Universiti Sains Malaysia}\\
{\normalsize 11800 USM Penang, Malaysia}}

\date{May 2005}
\maketitle

\begin{abstract}
We would like to present some exact SU(2) Yang-Mills-Higgs monopole solutions of half-integer topological charge. These solutions can be just an isolated half-monopole or a multimonopole with topological magnetic charge, $\frac{1}{2}m$, where $m$ is a natural number. These static monopole solutions satisfy the first order Bogomol'nyi equations. The axially symmetric one-half monopole gauge potentials possess a Dirac-like string singularity along the negative z-axis. The multimonopole gauge potentials are also singular along the z-axis and possess only mirror symmetries.   
\end{abstract}


\section{Introduction}
The SU(2) Yang-Mills-Higgs (YMH) field theory in $3+1$ dimensions, with the Higgs field in the adjoint representation possess both the magnetic monopole and multimonopole solutions \cite{kn:1}-\cite{kn:4}. The 't Hooft-Polyakov monopole solution belongs to the category of solutions which are invariant under a U(1) subgroup of the local SU(2) gauge group and has non zero Higgs mass and self-interaction. This numerical monopole solution of unit magnetic charge is spherically symmetric \cite{kn:1}. 

In general, configurations of the YMH field theory with a unit magnetic charge are spherically symmetric \cite{kn:1}-\cite{kn:3}. However this is not always true as we will show in this letter that unit magnetic monopole charge solutions need not possess only radial symmetry. In fact our unit magnetic charge solution posseses only mirror symmetry about a vertical plane through the z-axis. This particular solution is actually made up of two half-monopoles positioned at the origin. 
 
Multimonopole configurations with magnetic charges greater than unity possess at most axial symmetry \cite{kn:4} and it has been shown that these solutions cannot possess spherical symmetry \cite{kn:5}. 

Exact monopole and multimonopoles solutions exist in the Bogomol'nyi-Prasad-Sommerfield (BPS) limit \cite{kn:3}-\cite{kn:4}. Outside the BPS limit, when the Higgs field potential is non-vanishing only numerical solutions are known. Asymmetric multimonopole solutions with no rotational symmetry are also shown to exist \cite{kn:6}. However these solutions are numerical solutions even in the BPS limit.

Numerical axially symmetric monopoles-antimonopoles chain solutions which do not satisfy the Bogomol'nyi condition have also been discussed. These non-Bogomol'nyi solutions exist both in the limit of a vanishing Higgs potential as well as in the presence of a finite Higgs potential . Numerical BPS axially symmetric vortex rings solutions have also been reported  \cite{kn:7}.

Recently, we have also shown that the extended ansatz of Ref.\cite{kn:7}-\cite{kn:8} possesses more exact multimonopole-antimonopole configurations. We have also constructed the anti-configurations of all these multimonopole-antimonopole solutions \cite{kn:10} with all the magnetic monopole charges reversing their sign. Hence monopoles becomes antimonopoles and vice versa. We have also constructed exact magnetically neutral vortex rings in the presence of a antimonopole-monopole-antimonopole chain \cite{kn:11}.

However, all these monopole solutions reported so far are of integer topological monopole charges. It is our purpose in this letter to show the existence of half-integer topological monopole charge solutions. We have constructed a half-monopole configuration which is axially symmetric and possess a Dirac-like string singularity along the negative z-axis. 

However half-monopole solutions are not necessarily axially symmetric as we have also obtained a half-monopole solution that possesses only mirror symmetry about a vertical plane through the z-axis. In fact this half-monopole configuration is a member of a series of multimonopole solutions which we would like to label as the C solutions. The multimonopole of the C solutions possesses half-integer topological monopole charge.  

The existence of smooth Yang-Mills potentials which correspond to monopoles and vortices of one-half winding number has been demonstrated in Ref. \cite{kn:12} recently. However no exact or numerical solutions of the YMH equations have been given.

We briefly review the SU(2) Yang-Mills-Higgs field theory in the next section. We present the ansatz and some of its basic properties in section 3 and the axially symmetric half-monopole solution in section 4. In section 5, we discuss the half-integer topological charge multimonopole C solutions. We end with some comments in section 6.

\section{The SU(2) YMH Theory}
The SU(2) YMH Lagrangian in 3+1 dimensions with vanishing Higgs mass and self interaction is

\begin{equation}
{\cal L} = -\frac{1}{4}F^a_{\mu\nu} F^{a\mu\nu} + \frac{1}{2}D^\mu \Phi^a D_\mu \Phi^a.
\label{eq.1}
\end{equation}
The covariant derivative of the Higgs field and the gauge field strength tensor are given respectively by 
~$D_{\mu}\Phi^{a} = \partial_{\mu} \Phi^{a} + \epsilon^{abc} A^{b}_{\mu}\Phi^{c}$,~
and
~$F^a_{\mu\nu} = \partial_{\mu}A^a_\nu - \partial_{\nu}A^a_\mu + \epsilon^{abc}A^b_{\mu}A^c_\nu$,~
where $A^a_\mu$ is the gauge potential and the gauge field coupling constant $g$ is set to one without any loss of generality. The metric used is $g_{\mu\nu} = (-+++)$. The SU(2) internal group indices $a, b, c$ run from 1 to 3 and the spatial indices are $\mu, \nu, \alpha = 0, 1, 2$, and $3$ in Minkowski space.

The equations of motion that follow from the Lagrangian (\ref{eq.1}) are
\begin{eqnarray}
D^{\mu}F^a_{\mu\nu} - \epsilon^{abc}\Phi^{b}D_{\nu}\Phi^c = 0,~~~
D^{\mu}D_{\mu}\Phi^a = 0.
\label{eq.2}
\end{eqnarray}

The Bogomol'nyi equations is ~$B^a_i \pm D_i \Phi^a = 0$.~ 
The $\pm$ sign corresponds to monopoles and antimonopoles respectively for the usual BPS solutions \cite{kn:13}. In our case the multimonopole-antimonopoles of Ref. \cite{kn:9} are solved with the $+$ sign and the anti-multimonopole solutions are solvable with the $-$ sign \cite{kn:10}. 

The tensor to be identified with the electromagnetic field, as was proposed by 't Hooft \cite{kn:1} is
\begin{eqnarray}
F_{\mu\nu} &=& \hat{\Phi}^a F^a_{\mu\nu} - \epsilon^{abc}\hat{\Phi}^{a}D_{\mu}\hat{\Phi}^{b}D_{\nu}\hat{\Phi}^c\nonumber\\
	&=& \partial_{\mu}A_\nu - \partial_{\nu}A_\mu - \epsilon^{abc}\hat{\Phi}^{a}\partial_{\mu}\hat{\Phi}^{b}\partial_{\nu}\hat{\Phi}^c,
\label{eq.3}
\end{eqnarray}

\noindent where $A_\mu = \hat{\Phi}^{a}A^a_\mu$, the Higgs unit vector, $\hat{\Phi}^a = \Phi^a/|\Phi|$, and the Higgs field magnitude $|\Phi| = \sqrt{\Phi^{a}\Phi^{a}}$. 
The Abelian electric field is $E_i = F_{0i}$, and the Abelian magnetic field is $B_i = -\frac{1}{2}\epsilon_{ijk}F_{jk}$. 
The topological magnetic current \cite{kn:14} is defined to be 
\begin{eqnarray}
k_\mu = \frac{1}{8\pi}~\epsilon_{\mu\nu\rho\sigma}~\epsilon_{abc}~\partial^{\nu}\hat{\Phi}^{a}~\partial^{\rho}\hat{\Phi}^{b}~\partial^{\sigma}\hat{\Phi}^c,
\label{eq.4}
\end{eqnarray}

\noindent which is also the topological current density of the system and the corresponding conserved topological magnetic charge is
\begin{eqnarray}
M & = & \int d^{3}x~k_0 = \frac{1}{8\pi}\int \epsilon_{ijk}\epsilon^{abc}\partial_{i}\left(\hat{\Phi}^{a}\partial_{j}\hat{\Phi}^{b}\partial_{k}\hat{\Phi}^{c}\right)d^{3}x\nonumber\\
& = & \frac{1}{8\pi}\oint d^{2}\sigma_{i}\left(\epsilon_{ijk}\epsilon^{abc}\hat{\Phi}^{a}\partial_{j}\hat{\Phi}^{b}\partial_{k}\hat{\Phi}^{c}\right)\nonumber\\
& = & \frac{1}{4\pi} \oint d^{2}\sigma_{i}~B_i. 
\label{eq.5}
\end{eqnarray}

\section{The Magnetic Ansatz}
The magnetic ansatz of Ref. \cite{kn:8} is given by 
\begin{eqnarray}
A^a_0 &=& 0\nonumber\\
A_i^a &=& \frac{1}{r}\psi(r)\left(\hat{\theta}^{a}\hat{\phi}_i - \hat{\phi}^{a}\hat{\theta}_i\right)  + \frac{1}{r}R(\theta)\left(\hat{\phi}^{a}\hat{r}_i - \hat{r}^{a}\hat{\phi}_i\right)\nonumber\\
&+& \frac{1}{r}G(\theta,\phi)\left(\hat{r}^{a}\hat{\theta}_i - \hat{\theta}^{a}\hat{r}_i\right),\nonumber\\
\Phi^a &=& \Phi_{1}~\hat{r}^a + \Phi_{2}~\hat{\theta}^a + \Phi_3~\hat{\phi}^a,
\label{eq.6}
\end{eqnarray}

\noindent where $\Phi_1 = \frac{1}{r}\psi(r), ~\Phi_2 = \frac{1}{r}R(\theta), ~\Phi_3 = \frac{1}{r}G(\theta, \phi)$. The spherical coordinate orthonormal unit vectors, $\hat{r}_i, \hat{\phi}_i$, and $\hat{\theta}_i$ are defined by 
\begin{eqnarray}
\hat{r}_i &=& \sin\theta ~\cos \phi ~\delta_{i1} + \sin\theta ~\sin \phi ~\delta_{i2} + \cos\theta \delta_{i3},\nonumber\\
\hat{\theta}_i &=& \cos\theta ~\cos \phi ~\delta_{i1} + \cos\theta ~\sin \phi ~\delta_{i2} - \sin\theta ~\delta_{i3},\nonumber\\
\hat{\phi}_i &=& -\sin \phi ~\delta_{i1} + \cos \phi ~\delta_{i2}.
\label{eq.7}
\end{eqnarray}
The gauge fixing condition that we used here is the radiation or Coulomb gauge, $\partial^i A^a_i = 0$, $A^a_0 = 0$.

From the ansatz (\ref{eq.6}), the Abelian gauge potentials of Eq.(\ref{eq.3}) always vanish as $A_{\mu} = \hat{\Phi}^{a}A^{a}_{\mu} = 0$. Hence the Abelian electric field, $E_i$, is zero and the Abelian magnetic field, $B_i$, is independent of the gauge field potentials, $A^a_\mu$. To calculate for the Abelian magnetic field $B_i$, we rewrite the Higgs field of Eq.(\ref{eq.6}) from the spherical to the Cartesian coordinate system, \cite{kn:7}-\cite{kn:9} 
\begin{eqnarray}
\Phi^a &=& \Phi_{1}~\hat{r}^{a} + \Phi_{2}~\hat{\theta}^{a} + \Phi_3~\hat{\phi}^{a}\nonumber\\
&=& \tilde{\Phi}_1 ~\delta^{a1} + \tilde{\Phi}_2 ~\delta^{a2} + \tilde{\Phi}_3 ~\delta^{a3},
\label{eq.8}
\end{eqnarray}
\begin{eqnarray}
\mbox{where}~~~\tilde{\Phi}_1 &=& \sin\theta \cos \phi ~\Phi_1 + \cos\theta \cos \phi ~\Phi_2 - \sin \phi ~\Phi_3
= |\Phi|\cos\alpha \sin\beta\nonumber\\
\tilde{\Phi}_2 &=& \sin\theta \sin \phi ~\Phi_1 + \cos\theta \sin \phi ~\Phi_2 + \cos \phi ~\Phi_3
= |\Phi|\cos\alpha \cos\beta\nonumber\\
\tilde{\Phi}_3 &=& \cos\theta ~\Phi_1 - \sin\theta ~\Phi_2 = |\Phi|\sin\alpha.
\label{eq.9}
\end{eqnarray}
The Higgs unit vector is then simplified to 
\begin{eqnarray}
\hat{\Phi}^a &=& \cos\alpha \sin\beta ~\delta^{a1} + \cos\alpha \cos\beta ~\delta^{a2} + \sin\alpha ~\delta^{a3},\\
\label{eq.10}
\mbox{where},~~~\sin\alpha &=& \frac{\psi\cos\theta - R \sin\theta}{\sqrt{\psi^2+R^2+G^2}},\nonumber\\
\beta = \gamma - \phi,~~~\gamma &=& \tan^{-1}\left(\frac{\psi\sin\theta + R \cos\theta}{G}\right),
\label{eq.11}
\end{eqnarray}
and the Abelian magnetic field is found to be
\begin{eqnarray}
B_i &=& \frac{1}{r^2 \sin\theta}\left\{\frac{\partial\sin\alpha}{\partial\theta}\frac{\partial\beta}{\partial\phi} - \frac{\partial\sin\alpha}{\partial\phi}\frac{\partial\beta}{\partial\theta}\right\}\hat{r}_i\nonumber\\
&+& \frac{1}{r\sin\theta}\left\{\frac{\partial\sin\alpha}{\partial\phi}\frac{\partial\beta}{\partial r} - \frac{\partial\sin\alpha}{\partial r}\frac{\partial\beta}{\partial\phi}\right\}\hat{\theta}_i\nonumber\\
&+& \frac{1}{r}\left\{\frac{\partial\sin\alpha}{\partial r}\frac{\partial\beta}{\partial\theta} - \frac{\partial\sin\alpha}{\partial\theta}\frac{\partial\beta}{\partial r}\right\}\hat{\phi}_i.
\label{eq.12}
\end{eqnarray}
Defining the Abelian field magnetic flux as 
\begin{eqnarray}
\Omega = 4\pi M = \oint d^{2}\sigma_{i} B_i = \int B_{i}(r^{2}\sin\theta d\theta )\hat{r}_{i}~d\phi,
\label{eq.13}
\end{eqnarray}
the magnetic charge enclosed by the sphere center at $r=0$ and of fixed radius, $r_1$, is calculated to be, 
\begin{eqnarray}
M_{r_1} &=& \frac{1}{4\pi}\int^{2\pi}_0\int^\pi_0\left.\left(\frac{\partial\sin\alpha}{\partial\theta}\frac{\partial\beta}{\partial\phi} - \frac{\partial\sin\alpha}{\partial\phi}\frac{\partial\beta}
{\partial\theta}\right) d\theta d\phi\right|_{r=r_1}.
\label{eq.14}
\end{eqnarray}
Hence the magnetic charge enclosed by the sphere of infinite radius is denoted by $M_\infty$ and the magnetic charge enclosed by the sphere of vanishing radius, $r\rightarrow 0$, is denoted by $M_0$. 

To solve for solutions, the ansatz (\ref{eq.6}) is substituted into the equations of motion (\ref{eq.2}) as well as the Bogomol'nyi equations with the positive sign and these equations can be simplified to just four first order differential equations,
\begin{equation}
r\psi^{\prime} + \psi - \psi^2 = -p,
\label{eq.15}
\end{equation}
\begin{equation}
\dot{R} + R\cot\theta - R^2 = p - b^2\csc^2\theta,
\label{eq.16}
\end{equation}
\begin{equation}
\dot{G} + G\cot\theta = 0,~~~G^\phi\csc\theta - G^2 = b^2\csc^2\theta,
\label{eq.17}
\end{equation}
where $p$ and $b^2$ are arbitary constants. By letting $p=m(m+1)$, $m\geq -\frac{1}{2}$, the solutions to Eq.(\ref{eq.15}) and Eq.(\ref{eq.17}) are 
\begin{equation}
\psi=\frac{(m+1)-mr^{2m+1}}{1+r^{2m+1}}, ~~~G=b\csc\theta\tan(b\phi).
\label{eq.18}
\end{equation}
Hence $\psi(r)$ is a smooth and bounded function for all values of $r$ with boundary conditions $\psi(r)|_{r\rightarrow 0} = m+1$, and $\psi(r)|_{r\rightarrow \infty} = -m$. Therefore the magnitude and hence the vacuum expectation value of the Higgs field vanish as $1/r$ at large $r$. Since the solution, $\psi(r)$ does not vanish at $r=0$, the gauge potentials of Eq.(\ref{eq.6}) are singular at the origin and hence the energy of the system is not finite.  

For the profile function $G(\theta, \phi)$ to be a single value function, the parameter $b$ is restricted to only half-integer values.

\section{The Axially Symmetric Half-Monopole}
In order to obtain axially symmetric solutions, the parameter $b$, and hence the profile function $G(\theta,\phi)$ is set to zero to eliminate the $\phi$ dependence of the gauge potentials \cite{kn:11}. The axially symmetric half-monopole solution is obtained by put $m=-\frac{1}{2}$ into the simplified equations of motion and upon solving Eq.(\ref{eq.15}) and Eq.(\ref{eq.16}) the solution obtained is 
\begin{eqnarray}
\psi(r) = \frac{1}{2},~~~
R(\theta) = \frac{1}{2}\left\{\cot\theta-\frac{P_{\frac{1}{2}}(\cos\theta)}{P_{-\frac{1}{2}}(\cos\theta)}\csc\theta\right\}, 
\label{eq.19}
\end{eqnarray}
where $P_{\pm\frac{1}{2}}(\cos\theta)$ is the Legendre function of the first kind of degree $\pm\frac{1}{2}$. Hence the boundary conditions of the solution, Eq.(\ref{eq.19}), are ~$R(0) = 0,~~R(\pi) = -\infty$ and the gauge potentials possess a Dirac-like string singularity along the negative z-axis. 

Also $\sin\alpha = \frac{\psi\cos\theta - R \sin\theta}{\sqrt{\psi^2+R^2}}$ and $\beta = \pi/2 - \phi,$ are independent of distant $r$. Hence the topological magnetic charge does not depend on the radius of the enclosing sphere and 
\begin{eqnarray}
M = M_{\infty} = -\left.\frac{1}{2}\sin\alpha\right|^{\pi}_{0, r\rightarrow \infty} = \frac{1}{2}.
\label{eq.20}
\end{eqnarray}

There is also no zeros of the Higgs field in solution (\ref{eq.19}), and there exist only a half-monopole located at the origin, $r=0$, where the Higgs field is singular. The magnetic field, $B_i=B_r\hat{r}_i$, where $B_r=-\frac{1}{2}\frac{\partial}{\partial\theta}\sin\alpha$, of the half-monopole is axially symmetric about the z-axis and solely radial in direction. This is in contrast to the Dirac 1-monopole which is radially symmetrical. The magnetic field, $B_i$, points radially outwards from $\theta= 0 \dots 2.82$ radian, after which it changes sign and points inwards. It blows up along the the negative z-axis giving rise to the string singularity, Fig.(\ref{fig.1}). 

The existence of smooth Yang-Mills potentials which correspond to monopoles and vortices of one-half winding number in the SU(2) YMH field has also been demonstrated in Ref. \cite{kn:12}. The Abelian magnetic field of the half-monopole solution (\ref{eq.19}) is similar to that discussed in Ref. \cite{kn:12} in that the magnetic field is axially symmetrical and possess a Dirac-like string singularity along the negative z-axis, Fig.(\ref{fig.1}).   

The axially symmetric monopole solutions for $m > -\frac{1}{2}$, is discussed in a separate work \cite{kn:11} when $m$ is a natural number and in a later work when $m$ is a positive non-integer. 

\section{The C Series of Solutions}
The C solution is a series of multimonopole solutions with half-integer topological magnetic charge. The multimonopole is located at the origin, $r=0$, and has positive topological magnetic charge, $M= m \in \{0, \frac{1}{2}, 1, \frac{3}{2}, 2, \dots\}$. This series of solutions is solved by writing, $p=0$ and $b=m$ in Eq.(\ref{eq.15}) to Eq.(\ref{eq.17}). The solutions obtained are
\begin{eqnarray}
\psi(r)=\frac{1}{1+r},~~~R(\theta)= m\csc\theta,~~~G(\theta,\phi)=m\csc\theta \tan(m\phi).
\label{eq.21}
\end{eqnarray}
The boundary conditions are $\psi(r)|_{r\rightarrow 0}=1, ~\psi(r)|_{r\rightarrow \infty}=0$ and $G(\theta,0)=G(\theta,2\pi)=0$. 

The magnetic charge of the monopole at $r=0$ is calculated to be one-half of the normal t'Hooft-Polyakov monopole charge when $m=\frac{1}{2}$. This half-monopole solution of the C series is different from the half-monopole, Eq.(\ref{eq.19}), of the axially symmetric series in that it possesses only mirror symmetry at the vertical plane through the x-z axes. It however possesses similar string singularity along the negative z-axis. 

The magnetic field, $B_i=B_r\hat{r}_i + B_\theta\hat{\theta}_i + B_\phi\hat{\phi}_i$, of the C half-monopole is plotted Fig.(\ref{fig.2}). The radial magnetic field, $B_r$, of this half-monopole possesses a string singularity along the negative z-axis as shown by the plot of Fig.(\ref{fig.3}). 

When $m=0$ and 1, the C configurations possess topological magnetic charge one. The 1-monopole when, $m=0$, is just the radially symmetric Wu-Yang type 1-monopole \cite{kn:8}-\cite{kn:9} whereas the 1-monopole when, $m=1$,  possesses only mirror symmetry at the vertical x-z plane. The 3-D surface plot of the Abelian magnetic energy density, $B_iB_i$, reveals that this particular 1-monopole is actually made up of two half-monopoles.  

The C series of solutions continue to support higher topological magnetic charge multimonople as $m$ increases in steps of half. Hence when $m=\frac{3}{2}, 2, \frac{5}{2}, 3, \dots $, the topological charge of the multimonopole is $M=\frac{3}{2}, 2, \frac{5}{2}, 3, \dots $ respectively. All these C multimonopoles except for the case when, $m=0$, seem to be made up of half-monopoles.

\section{Comments}
We have presented two different half-integer topological magnetic charge monopole solutions. These two half-monopole solutions possess some similar descriptions to that of Ref. \cite{kn:12} in that they both possess a Dirac-like string singularity along the negative z-axis. The C half-monopole solution possesses only mirror symmetry at the vertical x-z plane. Hence half-monopole solutions can possess both axial symmetry, as well as mirror symmetry at a plane. Only axially symmetrical half-monopole Abelian magnetic field was discussed in Ref. \cite{kn:12}. 

All the C multimonopoles seem to be composed of $2m$ half-monopoles located at the origin when $m=\frac{1}{2}, 1, \frac{3}{2}, 2, \dots$. The C monopole configurations possess only mirror symmetries. 

Besides the half-integer topological magnetic charge solutions given here, there also exist other monopole series of solutions of similar property. These solutions are solved within the same magnetic ansatz of Eq.(\ref{eq.6}). The A2  multimonopole solutions of Ref. \cite{kn:9} and \cite{kn:10} possess topological magnetic charge, $M=\frac{5}{2}, 3, \frac{7}{2}, 4, \dots $, at the origin. There also exist other half-integer topological magnetic charge, axially symmetric solutions when, $-\frac{1}{2}<m<\frac{1}{2}$.  We will these cases in a later work.

The A1 multimonopole screening solutions \cite{kn:10} also possess multimonopole of half-integer topological charge, $M=\frac{5}{2}, 3, \frac{7}{2}, 4, \dots$, at the origin.

\section{Acknowlegements}
The author, Rosy Teh, would like to thank Universiti Sains Malaysia for the short term research grant (Account No: 304/PFIZIK/634039).

\newpage

\section*{FIGURE CAPTIONS}
\begin{figure}[tbh]
\vspace{5in}
\vskip1in
\hskip0.5in\special{bmp:Br_theta_AS_m=-0.5.jpg x=5in y=5in}
\caption{}
\label{fig.1}
\end{figure}

\begin{figure}[tbh]
\vspace{5in}
\hskip0.5in\special{bmp:Mag_field_C_m=0.5.jpg x=5in y=5in}
\caption{}
\label{fig.2}
\end{figure}

\begin{figure}[tbh]
\vspace{5in}
\hskip0.5in\special{bmp:Br_theta_C_m=0.5.jpg x=5in y=5in}
\caption{}
\label{fig.3}
\end{figure}

\begin{figure}[tbh]
\vspace{5in}
\hskip0.5in\special{bmp:energydensity_C_m=1.jpg x=5in y=5in}
\caption{.}
\label{fig.4}
\end{figure}

\end{document}